# Oxidation stability of confined linear carbon chains, carbon nanotubes, and graphene nanoribbons as 1D nanocarbons

Weili Cui[b], Takeshi Saito[c], Paola Ayala[b], Thomas Pichler[b], Lei Shi[a,b,*]

[a.] School of Materials Science & Engineering, Sun Yat-sen University, Guangzhou 510275, Guangdong, P. R. China.
[b.] University of Vienna, Faculty of Physics, 1090 Wien, Austria.
[c.] Nanomaterials Research Institute, National Institute of Advanced Industrial Science and Technology (AIST), Tsukuba, Ibaraki 305-8565, Japan.
*Corresponding Author. E-mail: shilei26@mail.sysu.edu.cn (Lei Shi).

Three typical one-dimensional (1D)/quasi-1D nanocarbons, linear carbon chains, carbon nanotubes, and graphene nanoribbons have been proven to grow inside single-wall carbon nanotubes. This gives rise to three types of hybrid materials whose behaviour and properties compared among each other are far from understood. After proving successful the synthesis of these nanostructured materials in recently published work, we have now been able to study their oxidation stability systematically by using resonance Raman spectroscopy. Surprisingly, the linear carbon chains, which have been theoretically predicted to be very unstable, are actually thermally stable up to 500 °C assisted by the protection of the carbon nanotube hosts. Besides, longer linear carbon chains inside narrower CNTs are more stable than the shorter ones inside larger tubes, suggesting that the thermal stability not only depends on the length of linear carbon chains alone, but it is correlated with the confinement of the host tubes in a more complicated manner. In addition, graphene nanoribbons come overall in view as the most stable confined structures. On the other hand, peculiarities like the higher stability of the (6,5) CNT over its (6,4) counterpart allow this study to provide a solid platform for further studies on the application of these 1D nanocarbons (including true 1D linear carbon chains) at ambient conditions.

## 1 Introduction

The discovery of carbon nanotubes (CNTs) has greatly motivated the studies on nanomaterials over the past three decades.[1] Especially, its one-dimensional (1D) structure results into a variety of outstanding mechanical, electrical, and optical properties.[2-5] Besides, the hollow cores of CNTs can make ideal templates to synthesize novel 1D materials,[6-8] such as: ultra-thin inner tubes,[9] graphene nanoribbons (GNRs),[10,11] linear carbon chains (LCCs),[12] ionic chains,[13] π-conjugated polymers,[14] sulphur chains,[15] metal chains,[16] among others. From the 1D nanocarbons, CNTs and GNRs have been largely studied in the past years, while LCCs are still under preliminary exploration. CNTs present superior electrical and thermal transport, thus they could be used as the ideal components for nanodevices and other applications.[17,18] On the other hand, GNRs are for instance promising as an alternative to copper integrated circuits as interconnect materials because of their planar structure, high conductivity, high thermal conductivity, and low noise.[19] As a true-1D nanocarbon, the stiffness, Young's modulus, specific area of the LCCs surpass that of the CNTs and GNRs from the theoretical point of view.[20] However, the experimental comparison of these three 1D nanocarbons, especially regarding their stability against oxidation, remains elusive.

The oxidation stability of the carbon allotropes can vary a lot. For example, graphite is stable up to 2000 °C,[21] whereas the burning temperature of CNTs is usually between 400-600 °C and it highly depends on the number of walls and the tube´s diameter.[22] For instance, single-wall CNTs (SWCNTs) can be opened, cut, and functionalized at around 400 °C.[23] Commonly, when the diameter of SWCNTs is less than 2nm, larger nanotube is more resistant to oxidation than the smaller one.[24] However, when nanotubes are inside other larger ones, they can technically be as stable as the outer larger tubes, since they can be protected by the external ones. For GNRs synthesized on a gold surface, it was found that the zigzag edges and armchair edges started to be oxidized at different temperatures, around 180 and 430 °C, respectively.[25] As for the LCCs, no specific experimental report on their thermal stability is available so far. However, it is predicted and expected that they are extremely reactive and therefore unstable against low temperature oxidation.[26,27] However, high-pressure experiments of LCCs inside CNTs indicate that the LCCs only can be destroyed above greater than 10 GPa, similar to the water chains or inner tubes inside SWCNTs, suggesting that the LCCs inside nanotubes can be much more stable than free carbon chains.[28-30] As mentioned above, comparison of the oxidation stability among the confined LCCs, nanotubes, and nanoribbons as 1D nanocarbons becomes highly demanded now.

In this work, we show a comparative study of the oxidation stability of LCCs, CNTs, and GNRs confined inside the SWCNTs by using resonance Raman spectroscopy as probe. Previously, we have successfully synthesized these ultra-thin inner CNTs,[31] GNRs,[32] and LCCs[33,34] by using the same batch of SWCNTs as templates. The innertubes and LCCs/GNRs studied in this work were synthesized simultaneously starting from the same original SWCNTs. Here we prove the difference in stability of these 1D nanocarbons as a function of temperature by monitoring the corresponding Raman response. This study makes evident that these three nanostructured materials are well protected by their host tubes, and among them, GNRs are overall the most stable. Although the true 1D LCCs is believed to be very reactive and unstable, the thermal stability of the confined LCCs is in general comparable to that of the other two quasi-1D nanocarbons, opening up the possibility for applications even in ambient conditions.

## 2 Results and discussion

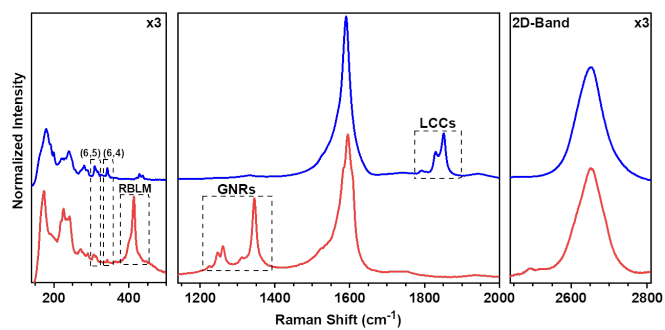

**Fig.1** Typical Raman spectra of linear carbon chains (upper one) and graphene nanoribbons (bottom one) formed inside carbon nanotubes (appear in both spectra) excited by a laser with wavelength of 568 nm.

Raman spectroscopy is a powerful tool to study the structure and properties of carbon nanomaterials, *e.g.* the Raman frequency of carbon nanomaterials depends on the type of the orbital hybridization of the atoms, from $sp^3$, $sp^2$ to $sp^1$.[35, 36] Carbon nanotubes and graphene nanoribbons with $sp^2$ hybridization normally show a very strong C-C bond stretching mode called G-band, which appears around 1600cm$^{-1}$, as seen in the spectra in Figure 1. Furthermore, a Raman mode known as radial breathing mode (RBM) is observable at low wavenumbers for nanotube diameters less than 2 nm, representing a signal arising from the carbon atoms moving in phase perpendicular to the tube axis. The RBM allows calculating the diameter and the chirality of the CNT.[37] For instance, (6, 5) and (6, 4) nanotubes can be identified as marked in the spectra (also see an enlarged RBM region in Figures 3a and 3b). Note that both the (6,5) and (6,4) tubes consist of several RBM components nearby, because each of them stays inside different outer tubes with varied interactions, resulting in small shifts of their RBM peaks. In addition to RBM, in Figure 1, the spectra of linear carbon chains (top spectra) and graphene nanoribbons (bottom spectra) hosted inside SWCNTs are also presented. For example, a radial breathing like mode (RBLM) for the GNRs, similar to the RBM of CNTs, is used as a fingerprint to determine their widths. The peak at 413 cm$^{-1}$, marked in Fig 1, is a signature of narrow ribbons less than 1 nm wide.[38] Additionally, the strong non-dispersive Raman lines at around 1260 and 1345 cm$^{-1}$ (the D-mode) correspond to the breathing-like vibrations of six-carbon-atom rings combined with C-H in plane bending.[32] Completing the identification of the type of structures, $sp^1$-hybridized nanocarbons also exhibit, a Raman mode between 1790 and 1860 cm$^{-1}$, which is labelled as LCC-mode and whose frequency is associated to the length of the LCC: *e.g.*, the peaks at 1790, 1800, and 1830 cm$^{-1}$ belong to long LCCs (LLCCs) with more than hundreds of carbon atoms, whereas the components at 1840 and 1850 cm$^{-1}$ correspond to relatively shorter LCCs with tens of carbon atoms.[39] Note that the LCCs were synthesized simultaneously with the inner tubes during the same annealing process. In order to analyze the oxidation process of the samples, the RBM mode of CNTs, the RBLM and D modes of GNRs, and the LCC-mode of LCCs were monitored with increased burning temperatures.

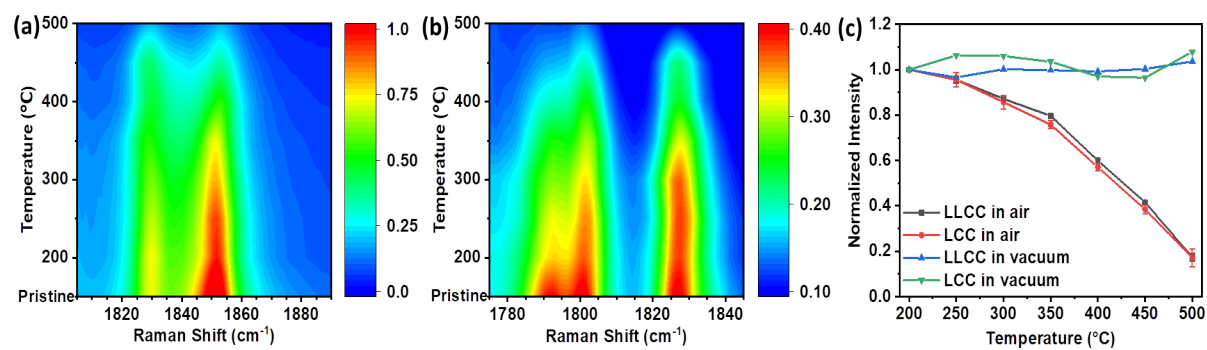

Fig. 2 Raman contour maps of LCCs with thermal oxidation at temperatures from 200 to 500 °C excited by (a) 568 and (b) 633 nm lasers. (c) Normalized relative intensity of LCC-band as a function of temperatures with oxidation in air or annealing in vacuum. The value of the error bar at each temperature is given by the intensity change after annealing in vacuum compared to the original intensity at 200 °C. Raman spectra of the samples annealed in vacuum are presented in Figures S2 and S3 in the supplementary information.

To gain a better insight, Figure 2 shows the short and long LCCs with band gaps around 2.25 and 1.85 eV measured by 568 and 633 nm laser wavelengths, which are respectively in resonance.[40] The Raman intensity of LCC-mode has been plotted into a contour map as a function of the oxidation temperature (The Raman spectra can be found in Figure S1 in the supplementary information). In general, the LCC-mode degrades with increased oxidation temperature. Surprisingly, both short and long LCCs continue to be presented even at temperatures up to 500 °C. Compared to unstable polyyne, a form of short carbon chains with alternated single-triple bonds,[41-43] the remarkable stability of the LCCs can be attributed to the presence of the CNTs as protector.[44] To quantify the change after each oxidation process, a relative intensity was calculated by normalizing the LCC-band intensity of the burned sample at each temperature to that of the sample treated at 200 °C. As shown in Figure 2c, the two categories for the shorter and longer chains behave no much difference under oxidation, therefore the LCCs with different lengths are all protected by the CNTs.

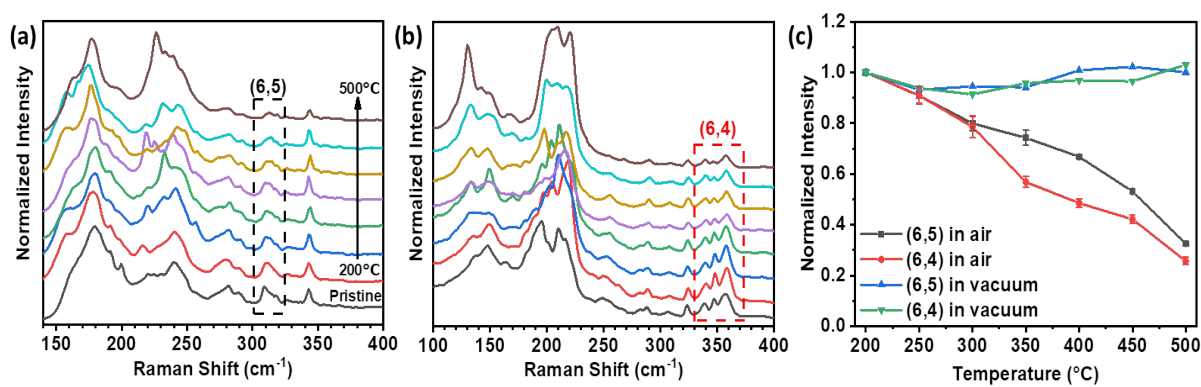

Fig.3 Raman spectra of the pristine and burned sample heated in air from 200 to 500 °C with a step of 50 °C excited by (a) 568 and (b) 633 nm lasers. (c) Normalized relative intensities of (6,5) and (6,4) inner tubes as a function of temperature with oxidation in air or annealing in vacuum. The value of the error bar at each temperature is given by the intensity change after annealing in vacuum compared to the original intensity at 200 °C. Raman spectra of the samples annealed in vacuum are presented in Figures S4 and S5 in the supplementary information.

Since the LCCs were synthesized simultaneously with the inner tubes during the same annealing process, this allows us comparing directly the oxidation stability between the inner tubes and the LCCs in the same sample. As shown in Figure 3, several peaks in the RBM region between 300 and 360 cm$^{-1}$ belong to the inner tubes with diameter from 0.65 to 0.85 nm, calculated by using the formula $D = 234/(\omega_{RBM} - 10)$, where D is the diameter of the nanotube in nm and $\omega_{RBM}$ is the Raman shift of RBM peaks in cm$^{-1}$ (wave numbers).[45] Also, their chiralities were assigned to (6,5) and (6,4) tubes according to "Kataura Plots".[46] Therefore, here we integrate all the peaks assigned to the same chirality to study the changes of the peak areas with the oxidation process. The relative intensity is plotted as a function of oxidation temperature in Figure 3c. Discernibly, the (6,5) tube is more resistant to the oxidation than the (6,4). Given its larger diameter, the (6,5) tube has less curvature compared to the (6,4) tube, and it is therefore more stable.[47] We have done the same for the GNRs, as illustrated in Figure 4. It is not surprising that all modes show almost the same behaviour and this is explained by the fact that they belong to the same GNRs. In other words, it is a hint that the method we have used here for comparison can be applied to study the oxidation stability of the materials.

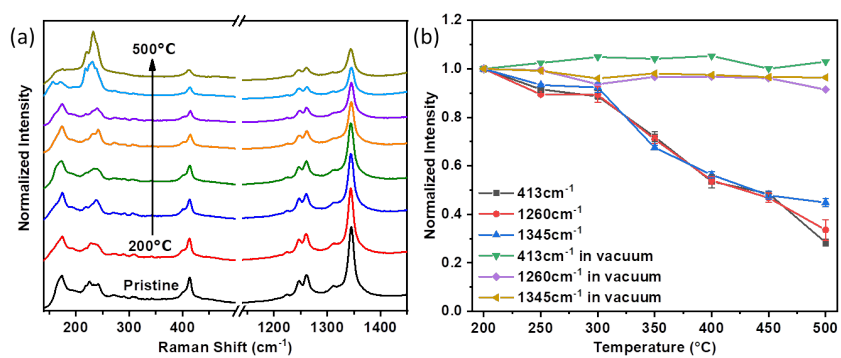

Fig. 4 Raman spectra of GNRs@SWCNTs before and after thermal oxidation treatment in air at temperatures from 200 to 500 °C with a step of 50 °C, excited by a 568 nm laser. (b) Normalized relative intensity of Raman peaks correspond to GNRs as a function of heating temperature. The value of the error bar at each temperature is given by the intensity change after annealing in vacuum compared to the original intensity at 200 °C. Raman spectra of the samples annealed in vacuum are presented in Figure S6 in the supplementary information.

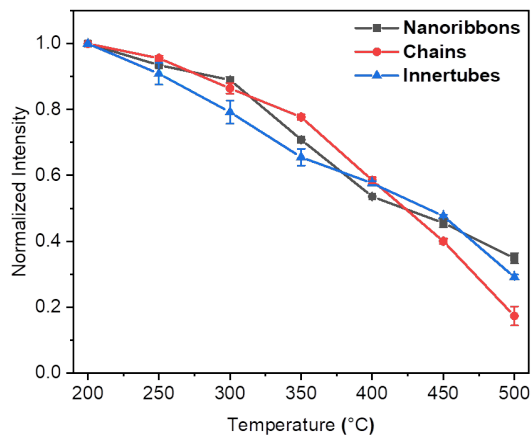

Fig. 5 Comparison of the oxidation stability of LCCs, GNRs, and inner CNTs. The value of the error bar at each temperature is given by the intensity change after annealing in vacuum compared to the original intensity at 200 °C.

To end with, we turn to the comparison of the oxidation stability of the LCCs, inner CNTs, and GNRs. Their normalized intensities versus oxidation temperatures are plotted together in Figure 5. The extremely subtle changes (in some cases negligible) attest for the efficiency of the external carbon nanotubes as protective layers of the different materials analysed. However, it has become clear that a GNR is more stable than an inner tube, even if its width is comparable to the diameter of the inner CNTs. This is a striking result because it hints that GNRs could potentially be synthesized from organic molecules inside SWCNTs as easily as the inner tubes can be made. [31] Indeed, recently we found that ferrocene molecules can be transformed more into GNRs than CNTs at the temperature below 800 °C. [48] And we believe this synthesis pathway can potentially turn into an effective method with other molecules as well. Finally, we emphasize that although the LCCs possess weaker oxidation resistance among the three 1D nanocarbons, still more than half of them can survive up to 400 °C, which makes them stable enough for most of the applications in future.

## 3 Conclusions

In this work, the oxidation stability of three 1D/quasi-1D nanocarbons, *i.e.* LCCs, CNTs, and GNRs, has been systematically studied. The Raman response of all these materials has been compared and the spectra have been weighed among each other. We found that they all can survive up to 500 °C while protected by SWCNTs as hosts. Among them, the GNRs are most stable, whereas the LCCs have a poorer performance but they are not defeated. In fact, LCCs still show superior stability than polyynes. At least half of the LCCs survive up to 400 °C, suggesting a bright future for the LCCs in applications.

## 4 Experimental

### 4.1 Sample preparation

SWCNTs prepared by the enhanced direct injection pyrolytic synthesis (eDIPS) method were used as templates for the synthesis of LCCs, GNRs, and ultra-thin inner CNTs. [49] The average diameter of the eDIPS SWCNTs is around 1.3 nm, which makes them to be suitable nanoreactors for the synthesis of inner CNTs/GNRs with diameter/width of ~0.7 nm. The description of the synthesis of these three nanocarbons can be found elsewhere. [32, 33, 50] In short, GNRs were obtained by transforming the ferrocene molecules inside SWCNTs in vacuum. [51] Inner CNTs were grown inside SWCNTs when annealing the SWCNTs at around 1460 °C in high vacuum ($10^{-6}$ mbar). In the same annealing process, LCCs were grown inside the new-formed inner tubes.

### 4.2 Oxidation process

In order to study the oxidation stability of these three 1D nanocarbons, the typical samples were put in an alumina crucible and heated simultaneously in an open furnace exposed to air from 200 to 500 °C with a step of 50 °C for 1h each time. The reference experiments for comparison were performed under high vacuum ($3 \times 10^{-6}$ mbar) instead of in air.

### 4.3 Characterization technique

The samples were measured at ambient conditions after each oxidation process by Raman spectroscope (Horiba, LabRAM HR800) with both 568 and 633 nm excitation laser wavelengths. The power of the laser on the sample was always kept at 0.5 mW to avoid any heating effect during the measurement. For ease of comparison, all the Raman spectra were normalized to the intensity of 2D-band in order to avoid any influence caused by the increased defects on the CNTs during the oxidation process.

## Conflicts of interest

There are no conflicts to declare.

## Acknowledgements

L. S. acknowledges the financial support by start-up funding from the Sun Yat-Sen University (29000-18841218). P. A. would like to acknowledge the contribution of the COST Action CA15107 (MultiComp). T. P. acknowledges the Austrian Science Funds (FWF, P27769-N20).